\documentclass[aps,twocolumn,showpacs,preprintnumbers,amsmath,amssymb,nofootinbib,superscriptaddress,showkeys]{revtex4-1}

\usepackage{hyperref}
\usepackage{float}
\usepackage{epsfig}
\usepackage{graphicx}
\usepackage{mathptmx}      % use Times fonts if available on your TeX system
\usepackage{latexsym}
\usepackage{natbib}
\begin{document}

\title{Phenomenological High Precision Neutron-Proton Delta-Shell Potential}

\author{R. Navarro P\'erez}\email{rnavarrop@ugr.es}
\affiliation{Departamento de F\'{\i}sica At\'omica, Molecular y
  Nuclear \\ and Instituto Carlos I de F{\'\i}sica Te\'orica y
  Computacional \\ Universidad de Granada, E-18071 Granada, Spain.}
\author{J.E. Amaro}\email{amaro@ugr.es} \affiliation{Departamento de
  F\'{\i}sica At\'omica, Molecular y Nuclear \\ and Instituto Carlos I
  de F{\'\i}sica Te\'orica y Computacional \\ Universidad de Granada,
  E-18071 Granada, Spain.}  \author{E. Ruiz
  Arriola}\email{earriola@ugr.es} \affiliation{Departamento de
  F\'{\i}sica At\'omica, Molecular y Nuclear \\ and Instituto Carlos I
  de F{\'\i}sica Te\'orica y Computacional \\ Universidad de Granada,
  E-18071 Granada, Spain.}

\date{\today}

\begin{abstract} 
\rule{0ex}{3ex} We provide a succesful fit for neutron-proton
scattering below pion production threshold up to LAB energies of $350
{\rm MeV}$.  We use seven high-quality fits based on potentials with
different forms as a measure of the systematic uncertainty.  We
represent the interaction as a sum of delta-shells in configuration
space below the $3 {\rm fm}$ and a charge dependent one pion exchange
potential above $3 {\rm fm}$ together with electromagnetic
effects. Special attention is payed to estimate the errors of the
phenomenological interaction.
\end{abstract}


\maketitle

%\section{Introduction}
%\label{sec:intro}

The study of the NN interaction has been a central and recurrent topic
in Nuclear Physics for many years (see
e.g.~\cite{Machleidt:1989tm,Machleidt:2011zz} and references
therein). The standard approach to constrain the interaction is to
analyze NN scattering data below pion production threshold and to
undertake a partial wave analysis (PWA), the quality of the fit being
given by the $\chi^2 / {\rm d.o.f}$ value. Only by the mid 90's was it
possible to fit about 4000 selected NN scattering data after
discarding about further 1000 of $3\sigma$ inconsistent data with a
$\chi^2 /{\rm d.o.f} \lesssim 1 $ and incorporating charge dependence
(CD) for the One Pion Exchange (OPE) potential as well as magnetic and
vacuum polarization effects~\cite{Stoks:1993tb}.  This benchmark
partial wave analysis (PWA) was carried out using an energy dependent
potential for the short range part for which nuclear structure
calculations become hard to formulate. Thus, energy independent {\it
  high quality} potentials were subsequently produced with almost
identical $\chi^2/{\rm d.o.f} \sim 1$ for a gradually increasing
database~\cite{Stoks:1994wp,Wiringa:1994wb,Machleidt:2000ge,Gross:2008ps}.
While any of these potentials provides individually satisfactory fits
to the available data there are no published error estimates of the
potential parameters. Moreover it should also be noticed that the
existing high-quality potentials are different in their specific form;
they range from local to non-local in different versions of
nonlocality.  Thus, scattering phase-shifts and observable amplitudes
are not identical and in fact the existing set of high quality
potentials as a whole provides a distribution of scattering
observables accounting for systematic uncertainties in addition to the
statistical uncertainties obtained from the fitted data for each
individual potential. Given the fact that these interactions are just
contrained to the elastic scattering data (and eventually to the
deuteron) which go up to the pion production threshold, one is
physically probing the interaction with a resolution not finer than
the shortest de Broglie wavelength $\Delta \lambda=\hbar / \sqrt{M_N
  m_\pi} \sim 0.5 {\rm fm} $. Thus, for practical purposes it may be
advantageous to consider {\it coarse grained}
interactions~\cite{Perez:2011fm}. This is actually the physics
underlying the so-called $V_{\rm low k}$ approach~\cite{Bogner:2003wn}
in which an effective interaction in a restricted model space is
built.  By starting from {\it different} high-quality potentials with
a common charge dependent OPE interaction, the CM-momenta above
$\Lambda \sim \sqrt{M_N m_\pi} $ are eliminated by a suitable
transformation and a remarkable universal interaction is obtained for
$p \le \Lambda$.  Many of the applications of such an appealing
interaction have recently been reviewed~\cite{Bogner:2009bt}.

\begin{figure*}[ht]
\begin{center}
\includegraphics[width=\textwidth]{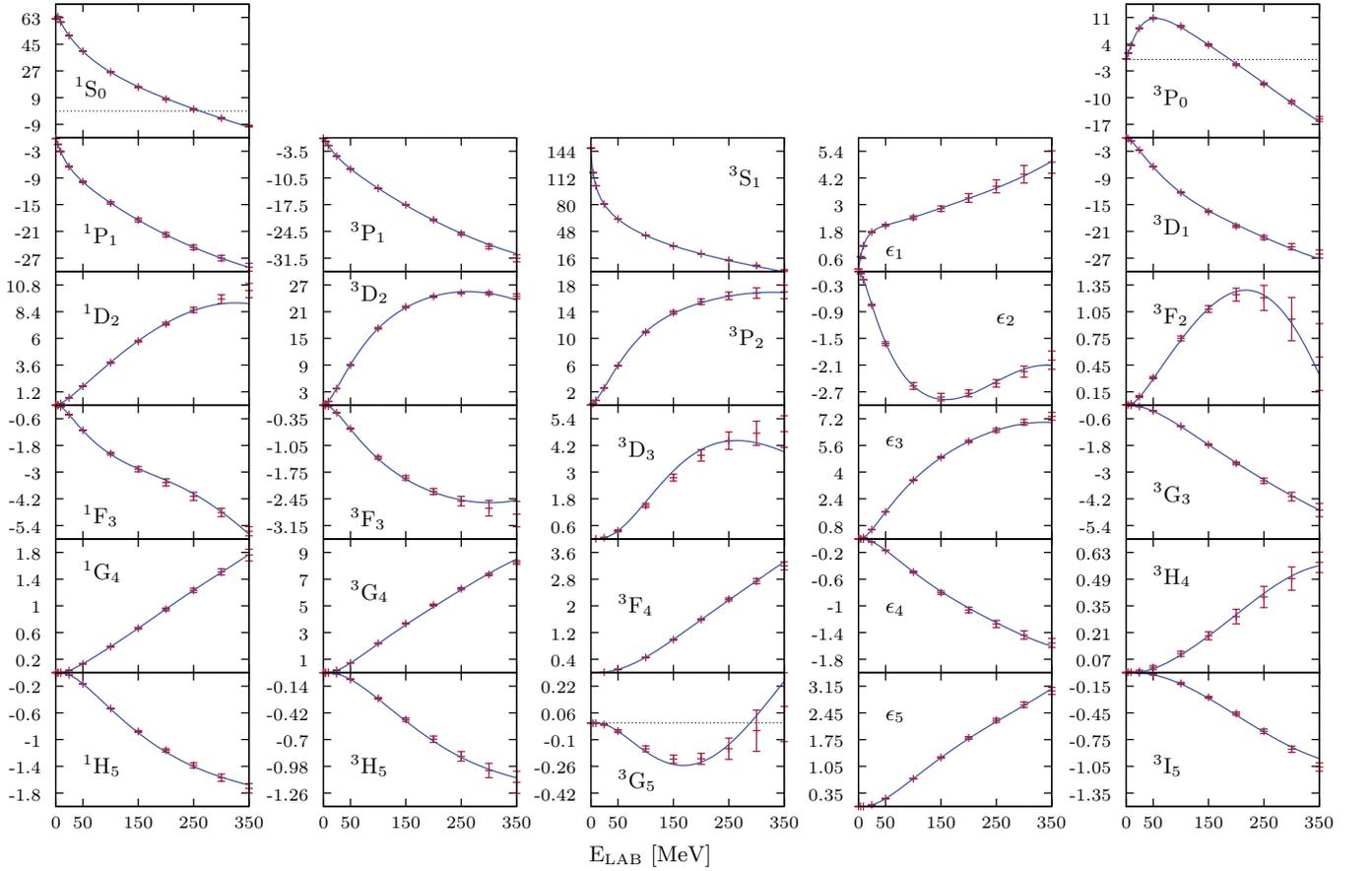} 
\end{center}
\caption{np phase shifts in degrees with $J\leq 5$ as a
function of the LAB energy in MeV. The solid line is calculated with the
fitted potential and the points with error bars represent the mean value
and standard deviation of the compilation of the PWA~\cite{Stoks:1993tb} and the six high quality
  potentials~\cite{Stoks:1994wp,Wiringa:1994wb,Machleidt:2000ge,Gross:2008ps}. We fit the energies $E_{\rm LAB} =
1,5,10,25,50,100,150,200,250,300,350$ MeV.}
\label{fig:all-phases}
\end{figure*}

\begin{table*}[htb]
	\centering
      \caption{Fitting delta-shell parameters $(\lambda_n)^{JS}_{l,l'}
        $ (in ${\rm fm}^{-1}$) with their errors for all states in the
        $JS$ channel and the corresponding $\chi^2$-value for $J \leq
        5$ in np scattering. We take $N=5$ equidistant points with
        $\Delta r = 0.6$fm. $-$ indicates that the corresponding
        $(\lambda_n)^{JS}_{l,l'}
        =0$. }
      \label{tab:Fits}
	\begin{tabular*}{\textwidth}{@{\extracolsep{\fill}}c c c c c c c c}
            \hline
            \hline\noalign{\smallskip}
		Wave  & $\lambda_1$ & $\lambda_2$ & $\lambda_3$ & $\lambda_4$ & $\lambda_5$ & $\chi^2/$D.o.F \\
%		  & & fm & fm & fm$^{-1}$ & fm$^{-1}$ & fm$^{-1}$ &  \\
            \hline\noalign{\smallskip}

            $^1S_0$  &  2.12(7) & -0.987(7)&   $-$    & -0.087(2)&   $-$    &  0.3476  \\
            $^3P_0$  &   $-$    &  1.26(4) & -0.43(1) &   $-$    & -0.037(2)&  0.6589  \\
            $^1P_1$  &   $-$    &  1.23(2) &   $-$    &  0.079(4)&   $-$    &  0.0088  \\
            $^3P_1$  &   $-$    &  1.33(2) &   $-$    &  0.053(2)&   $-$    &  0.4323  \\
            $^1D_2$  &   $-$    &   $-$    & -0.252(3)&   $-$    &-0.0163(9)&  0.6946  \\
            $^3D_2$  &   $-$    &   $-$    & -0.596(8)& -0.08(1) & -0.050(4)&  0.6144  \\
            $^1F_3$  &   $-$    &   $-$    &  0.34(1) &   $-$    &  0.010(2)&  0.3812  \\
            $^3F_3$  &   $-$    &   $-$    &   $-$    &  0.060(2)&   $-$    &  0.4177  \\
            $^1G_4$  &   $-$    &   $-$    & -0.22(2) &   $-$    &-0.0137(9)&  0.8090  \\
            $^3G_4$  &   $-$    &   $-$    &   $-$    & -0.267(3)&   $-$    &  1.8670  \\
            $^1H_5$  &   $-$    &   $-$    &   $-$    &  0.071(8)&   $-$    &  0.6577  \\
            $^3H_5$  &   $-$    &   $-$    &   $-$    &  0.04(1) &  0.0000  &  0.4193  \\

            $^3S_1$          &  1.57(4) & -0.40(1) &   $-$    & -0.064(3)&   $-$    &  \\
            $\varepsilon_1$  &   $-$    & -1.69(1) & -0.379(4)& -0.216(5)& -0.027(3)&  \\
            $^3D_1$          &   $-$    &   $-$    &  0.52(2) &   $-$    &  0.041(3)&  0.4313  \\
            $^3P_2$          &   $-$    & -0.415(6)&   $-$    &-0.0384(9)&   $-$    &  \\
            $\varepsilon_2$  &   $-$    &  0.65(1) &   $-$    &  0.106(2)&   $-$    &  \\
            $^3F_2$          &   $-$    &   $-$    &  0.14(3) & -0.076(6)&   $-$    &  0.3881  \\
            $^3D_3$          &   $-$    &   $-$    &   $-$    &   $-$    &   $-$    &  \\
            $\varepsilon_3$  &   $-$    &   $-$    & -0.47(3) & -0.24(1) & -0.020(4)&  \\
            $^3G_3$          &   $-$    &   $-$    &   $-$    &  0.101(6)&   $-$    &  0.6806  \\
            $^3F_4$          &   $-$    &   $-$    & -0.163(4)&   $-$    &-0.0101(4)&  \\
            $\varepsilon_4$  &   $-$    &   $-$    &   $-$    &  0.108(3)&   $-$    &  \\
            $^3H_4$          &   $-$    &   $-$    &   $-$    &   $-$    & -0.010(1)&  0.2659  \\
            $^3G_5$          &   $-$    &   $-$    &   $-$    &  0.025(4)&   $-$    &  \\
            $\varepsilon_5$  &   $-$    &   $-$    &   $-$    & -0.35(1) &   $-$    &  \\
            $^3I_5$          &   $-$    &   $-$    &   $-$    &   $-$    &   $-$    &  0.5354  \\
            \noalign{\smallskip}\hline
            \hline
	\end{tabular*}
\end{table*}

On the other hand, when switching from the NN problem to the many body
nuclear problem the features and the form of the interaction are
relevant in terms of computational cost and feasibility (see
e.g.~\cite{Pieper:2001mp} and references therein). The lack of
knowledge of a precise potential form with finer resolution than
$\Delta r \sim 0.5 {\rm fm}$ suggests to search for a description of
scattering data directly in terms of a coarse grained potential
sampled at some sensible ``thick points''. Any sampling procedure
necessarily redistributes the interaction strength and smoothes the
potential as compared to the zero resolution limit $\Delta r \to 0$
implicit in most potential approaches and generating the troublesome
short distance cores.  This requires short distance correlations in
the wave function to ensure the finiteness of the
energy~\cite{Pieper:2001mp}. A desirable way to sample the interaction
is to provide an acceptable $\chi^2$-fit with the minimal number of
sampling points~\cite{taylor-book}; by implementing this minimal
sampling we just try to avoid statistical dependence between the
strenghts of the potential at the chosen sampling points. Our
motivation to proceed in this fashion is to make error propagation in
nuclear structure calculations more direct since in the absence of
statistical correlations errors may just be added in
quadrature~\cite{NavarroPerez:2012vr}.

In the present work we provide another high-quality potential
accommodating these desirable features and (unlike the previous
approaches) we undertake an analysis of the systematic errors.  We do
this by assuming that the systematic error inherent to any specific
choice of the potential form corresponds to individual uncorrelated
measurements~\footnote{We have actually checked that, within the
  corresponding statistical uncertainty, the absense of correlations
  among the different partial waves of the PWA and the six high
  quality potentials by a direct evaluation of the correlation
  coefficient (see Ref.~\cite{taylor-book}for a definition) holds for
  every single LAB energy below $350$MeV.}. Hence we may invoke the
central limit theorem to undertake the traditional statistical
treatment to the mean average and the corresponding standard deviation
of
Refs.~\cite{Stoks:1993tb,Stoks:1994wp,Wiringa:1994wb,Machleidt:2000ge,Gross:2008ps}
without any further ado. We will use this compilation as our database.

A convenient representation to sample the the short distance component
of the NN interaction was already suggested by
Aviles~\cite{Aviles:1973ee} almost 40 years ago in terms of
delta-shells which for any
partial wave $^{2S}(l',l)_J$ we take as 
\begin{eqnarray}
V^{JS}_{l,l'}(r) = \frac{1}{2\mu_{np}}\sum_{n=1}^N  (\lambda_n)^{JS}_{l,l'} \delta(r-r_n)  \qquad r \le r_c
\label{eq:ds-pot}
\end{eqnarray}
with $\mu_{np}$ the reduced np-mass and $r_c$ to be specified
below. In the spirit of
Refs.~\cite{Stoks:1993tb,Stoks:1994wp,Wiringa:1994wb,Machleidt:2000ge,Gross:2008ps},
for $r \ge r_c $ we use the well known 
long-distance tail of the NN potential
\begin{equation}
V(\vec{r}) = V_{EM}(\vec{r})+V_{OPE}(\vec{r}), \qquad r > r_c,
\end{equation}
where $V_{EM}$ is the electromagnetic 
potential of Ref.~\cite{Wiringa:1994wb},
and $V_{OPE}$ is the one-pion-exchange potential.

The solution of the corresponding Schr\"odinger equation in coupled
channels is straightforward; for any $r_n < r < r_{n+1} $ with $r_N <
r_c$ we have free particle solutions and log-derivatives are
discontinuous at the $r_n$-points so that one generates an accumulated
S-matrix at any sampling point providing a discrete version of
Calogero's variable phase equation~\cite{Calogero:1965}.  Although
this potential is formally local, the fact that we are coarse graining
the interaction enables to encode efficiently nonlocalities operating
below the finest resolution $\Delta r$. Of course, once we admit that
the interaction below $r_c$ is unknown there is no advantage in
prolonging the well-known charge-dependent OPE tail and other
electromagnetic effects for $r< r_c$.  The low energy expansion of the
discrete variable phase equations was used in
Ref.~\cite{PavonValderrama:2005ku} to determine threshold parameters
in all partial waves. The relation to the well-known Nyquist theorem
of sampling a signal with a given bandwidth has been discussed in
Ref.~\cite{Entem:2007jg}.  Some of the advantages of using this simple
potential for Nuclear Structure calculations as well as the connection
to the $V_{\rm low k}$ approach have been spelled out
already~\cite{Perez:2011fm}.

\begin{table*}[htb]
	\centering
	\caption{Deuteron static properties compared with empirical values and high-quality potentials calculations}
	\label{tab:DeuteronP}
	\begin{tabular*}{\textwidth}{@{\extracolsep{\fill}}l l l l l l l l }
      \hline
      \hline
            & Delta Shell & Empirical\cite{VanDerLeun1982261,Borbély198517,Rodning:1990zz,Klarsfeld1986373,Bishop:1979zz,deSwart:1995ui} & Nijm I~\cite{Stoks:1994wp}   & Nijm II~\cite{Stoks:1994wp}  & Reid93~\cite{Stoks:1994wp}   & AV18~\cite{Wiringa:1994wb} & CD-Bonn~\cite{Machleidt:2000ge}  \\
% & Spect~\cite{Gross:2008ps} 0.0256(4) 0.8777(15) missing others
      \hline
		$E_d$(MeV)              & 2.2(2)    & 2.224575(9)    & Input    & Input    & Input    & Input  & Input           \\
		$\eta$                  & 0.025(2)  & 0.0256(5)      & 0.02534  & 0.02521  & 0.02514  & 0.0250 & 0.0256            \\
		$A_S ({\rm fm}^{1/2})$  & 0.88(3)   & 0.8781(44)     & 0.8841   & 0.8845   & 0.8853   & 0.8850  & 0.8846    \\
		$r_m ({\rm fm})$        & 1.97(8)   & 1.953(3)       & 1.9666   & 1.9675   & 1.9686   & 1.967 &  1.966           \\
		$Q_D ({\rm fm}^{2}) $   & 0.272(9)  & 0.2859(3)      & 0.2719   & 0.2707   & 0.2703   & 0.270  & 0.270   \\
		$P_D$                   & 5.7(2)    & 5.67(4)        & 5.664    & 5.635    & 5.699    & 5.76  & 4.85              \\
		$\langle r^{-1} \rangle ({\rm fm}^{-1})$& 0.45(1)   &                &          & 0.4502   & 0.4515   &  & \\
      \hline \hline
	\end{tabular*}
\end{table*}

\begin{figure*}[ht]
\begin{center}
\includegraphics[width=\textwidth]{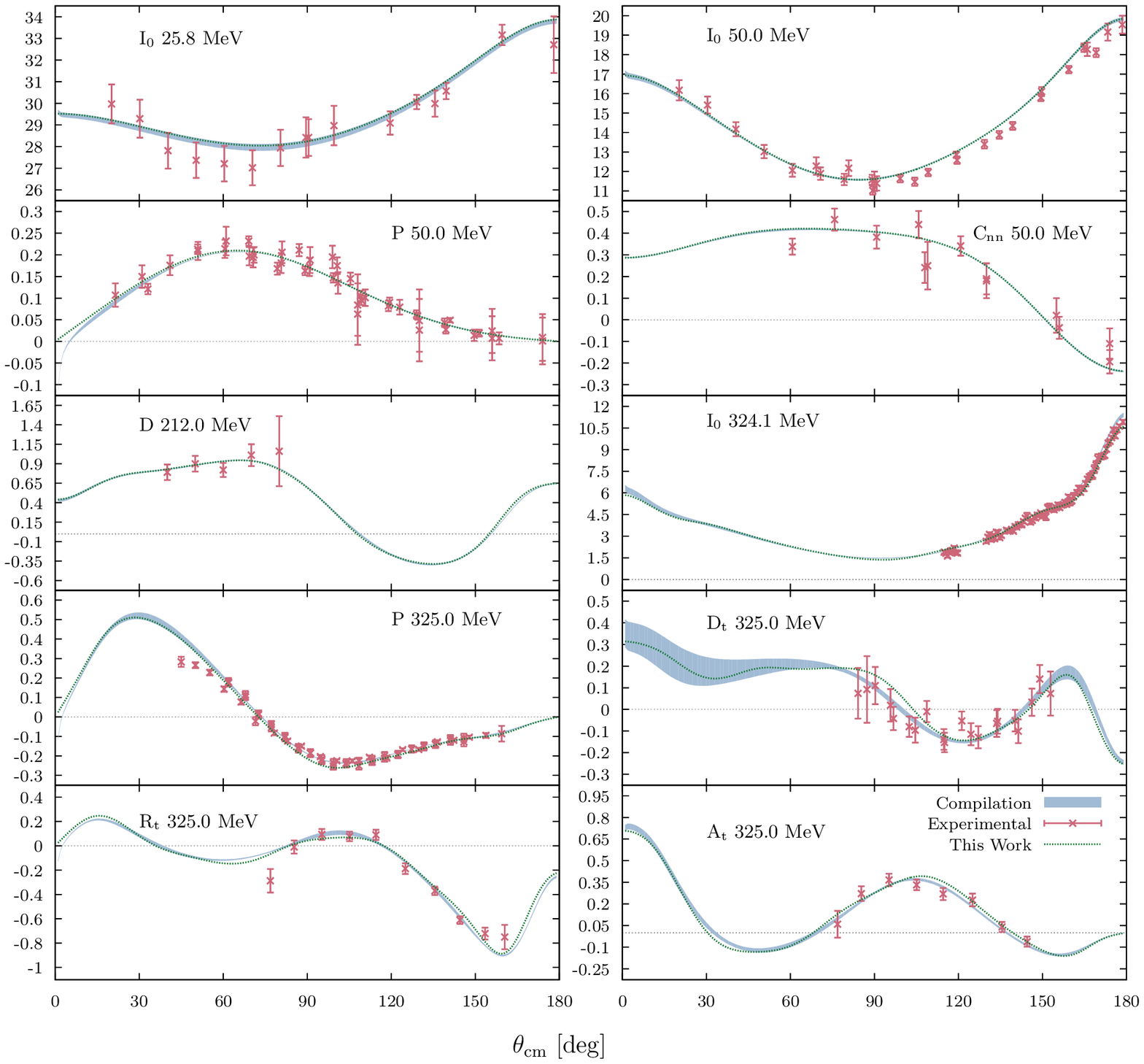} 
\end{center}
\caption{Some np scattering observables for several energies in the
  laboratory system as a function of the CM angle.  The short-dashed
  line denotes the predictions by our delta-shell model. The band
  represents the compilation the six high quality
  potentials~\cite{Stoks:1993tb,Stoks:1994wp,Wiringa:1994wb,Machleidt:2000ge,Gross:2008ps}
  which provided a $\chi^2/ {d.o.f} \lesssim 1 $.  For references for
  the experimental data see \url{http://nn-online.org} and
  \url{http://gwdac.phys.gwu.edu/} . The notation is as follows: $I_o $
  differential cross section, $P$ polarization, $D$ depolarization,
  $R$ rotation parameter, $A_t$,, $D_t$, and $R_t$, polarization transfer
  parameters, $C_{nn}$ spin correlation parameter.  For notation and
  further explanations see
  Refs.~\cite{bystricky:jpa-00208735,Lafrance80}}
\label{fig:observables}
\end{figure*}

We use the LAB energy values usually listed in the
high-quality potentials, namely $E_{\rm
  LAB}=1,5,10,25,50,100,150,200,250,300,350\,{\rm MeV}$ and fit to the
mean phase-shift values at those energies with an error equal to the standard
deviation. This energy grid is sufficiently dense to 
account for the systematic errors due to the
different representations of the
potentials~\cite{Stoks:1993tb,Stoks:1994wp,Wiringa:1994wb,Machleidt:2000ge,Gross:2008ps}. We
find that they are generally larger than those quoted by the original
PWA where only statistical uncertainties where explicitly discussed
for a {\it fixed} potential form~\cite{Stoks:1993tb}. With these data
sets and the given energies we undertake a phase-shift fit and
determine errors using the standard covariance matrix.

%As mentioned
%the discrepancies among the different high-quality potentials are
%generally larger than the errors quoted in the original PWA, thus the
%systematic error due to the form difference in the potential dominates
%the uncertainties.

As expected from Nyquist sampling theorem, we need at most $N=5$
sampling points which for simplicity are taken to be equidistant with
$\Delta r = 0.6 {\rm fm}$ between the origin and $r_c=3 {\rm
  fm}$. This is the minimal number which provides an acceptable fit to
the data compiled from
Refs.~\cite{Stoks:1993tb,Stoks:1994wp,Wiringa:1994wb,Machleidt:2000ge,Gross:2008ps}. Our
results for the np phase-shifts for all partial waves with total
angular momentum $J \le 5$ are depicted in
Fig.~\ref{fig:all-phases}. The fitting parameters
$(\lambda_n)^{JS}_{l,l'} $ entering the delta-shell potentials,
Eq.~(\ref{eq:ds-pot}), are listed in Table~\ref{tab:Fits} with their
deduced uncertainties.  Of course, a definitive assesment on
systematic errors would require testing {\it all possible} potential
forms. Thus, the errors will generally be larger than those estimated
here.  We find that correlations among the different
$(\lambda_n)^{JS}_{l,l'} $ values within a given partial wave channel
are unimportant, and hence these parameters are essentially
independent from each other. This is a direct consequence of our
strategy to minimize the number of sampling points. We find that
introducing more points or equivalently reducing $\Delta r$ generates
unnecessary correlations. Also, lowering the value of $r_c$ below $3
{\rm fm}$, requires overlapping the short-distance potential,
Eq.~(\ref{eq:ds-pot}), with the OPE plus em corrections.

We determine the deuteron properties by solving the bound state
problem in the $^3S_1-^3D_1$ channel using the corresponding
parameters listed in Table~\ref{tab:Fits}. The
predictions are presented in table~\ref{tab:DeuteronP} where our
quoted errors are obtained from propagating
Table~\ref{tab:Fits}. The comparison with experimental
values or high quality potentials where the binding energies are used
as an input is satisfactory. This is partly due to the fact that
theoretical errors are about $10 \%$. Of course, one may improve on
this by using the deuteron binding energy as an input as in
Refs.~\cite{Stoks:1993tb,Stoks:1994wp,Wiringa:1994wb,Machleidt:2000ge,Gross:2008ps}. 
% and other deuteron properties
%increasing the accuracy of the $^3S_1-^3D_1$ channel parameters, but
%not affecting the accuracy in other channels.

Fitting to phase-shifts to some accuracy does not necessarily provide
angle dependent scattering amplitudes to the same accuracy because
errors are finite and the relation between phase shifts and
observables is non-linear. This is often the case when the form of the
potential is kept fixed, so that the channel by channel fit is usually
taken as a first step which is afterwards refined by a full fledged
analysis of differential or scattering observables and polarization
data~\cite{Stoks:1993tb,Stoks:1994wp,Wiringa:1994wb,Machleidt:2000ge,Gross:2008ps}. To
check our strategy of fitting first phase-shifts and determine
observables afterwards we proceed as follows. The complete on-shell np
scattering amplitude contains five independent complex quantities,
which we choose for definiteness as the Wolfenstein parameters and
denote generically as a 10-dimensional array $(a_1, \dots, a_{10}) $,
which could be determined directly from experiment as shown in
Ref.~\cite{bystricky:jpa-00208735,Lafrance80} (see also
~\cite{springerlink:10.1007/s00601-010-0205-6} for an exact analytical
inversion). We follow an alternative procedure and construct, out of
the high-quality analyses, the corresponding mean value of the
Wolfenstein parameters,$\bar a_i(E_{\rm LAB},\theta)$, with their
corresponding standard deviations, $\Delta a_i(E_{\rm LAB},\theta)$,
for any given LAB energy and scattering angle $\theta$ as
\begin{eqnarray}
\chi^2 (E_{\rm LAB},\theta)  = \sum_{i=1}^{10} \left[\frac{\bar a_i(E_{\rm LAB},\theta) - a_i(E_{\rm LAB},\theta)}{\Delta a_i(E_{\rm LAB},\theta)} \right]^2
\end{eqnarray}
The total $\chi^2$ value is obtained as an average over the chosen
reference energies $E_{\rm
  LAB}=1,5,10,25,50,100,150,200,250,300,350\,{\rm MeV}$ and a dense
sampling of $\theta$-values~\footnote{A finer energy grid
for the $\chi^2$/d.o.f. remains stable and below unity.}. The result
is
$$
\chi^2 / {\rm d.o.f.} =  0.78 \,  
$$ which is equivalent to carry a complete $\chi^2$-fit to the mean
average scattering amplitude. In Fig.~\ref{fig:observables} we
illustrate the situation for a set of observables as a function of the
CM angle and for several energies. One sees that our model 1)
describes the data within experimental uncertainties and 2) it mostly
agrees with the six high quality
potentials~\cite{Stoks:1993tb,Stoks:1994wp,Wiringa:1994wb,Machleidt:2000ge,Gross:2008ps}
except marginally when no data are available.

To summarize, we have determined a high-quality neutron-proton
interaction which is based on a few delta-shells for the lowest
partial waves in addition to charge-dependent electromagnetic
interactions and one pion exchange and provides a good starting point
for Nuclear Physics applications. We provide error estimates on our
fitting parameters accounting both for systematic and statistical
uncertainties of present day high-quality analyses of neutron-proton
scattering data.  Deuteron properties are compatible with experimental
or reccomended values.  Our method allows to coarse grain long-range Coulomb
interactions in a rather natural way and hence to discuss
proton-proton scattering data.

{\it We want to thank D.R. Entem and M. Pav\'on Valderrama for a
  critical reading of the manuscript and Franz Gross for
  communications. This work is supported by Spanish DGI (grant
  FIS2011-24149) and Junta de Andaluc{\'{\i}a} (grant FQM225).
  R.N.P. is supported by a Mexican CONACYT grant.}

%\bibliography{high_q_fit}
%Merlin.mbs v4.21 2009-07-09.
%

\end{document}